\documentclass[preprint,12pt]{elsarticle}

\usepackage{amssymb}
\usepackage{float}

\begin{document}

\begin{frontmatter}



\title{Results and prospects of deep under-ground, under-water and
  under-ice experiments}

%%\author[label1,label2]{<author name>}
%%\address[label1]{<address>}

\author[IFIC]{J. D. Zornoza}
\ead{zornoza@ific.uv.es}
\address[IFIC]{IFIC, Instituto de F\'{i}sica Corpuscular (CSIC-Universidad de
Valencia, Ed. Institutos de Investigaci\'{o}n, AC22085, E46071, Valencia, Spain}

\begin{abstract}

  Astroparticle experiments have provided a long list of achievements
  both for particle physics and astrophysics. Many of these
  experiments require to be protected from the background produced by
  cosmic rays in the atmosphere. The main options for such protection
  are to build detectors deep under ground (mines, tunnels) or in the
  deep sea or antarctic ice. In this proceeding we review the main
  results shown in the RICAP 2013 conference related with these kind
  of experiments and the prospects for the future.

\end{abstract}


\end{frontmatter}

%%\linenumbers

\section{Introduction}
\label{intro}

Astroparticle experiments have provided richful physics results for
many years. In this conference have seen many examples of the most
recent advances and the prospects for the future, showing that this
yield is growing and will continue giving us answers (and new
questions) during the the following years. 

In this paper I will report on the results presented which are related
with one three following topics: dark matter, under-ground experiments
and neutrinos.

\section{Dark matter}
\label{dm}

Evidence for dark matter has been accumulating for almost one
century~\cite{theory1}. The experimental hints for its existence include galaxy
clusters, the rotation curves of galaxies, structure formation,
filaments, the Bullet cluster, etc. The main ingredient of the
Universe is dark energy. From the last results of Planck
satellite, the dark energy content has been estimated
in 68.3\%. Dark matter contribution is 26.8\% and ordinary matter is
just 4.9\% of the total. In other words, approximately 85\% of the
matter in the Universe is dark matter. The basic requirements for a
good dark matter candidate are to be stable (or very long-lived), neutral and with an
interaction cross section of the order of the one of the weak interaction. The
only viable candidate within the Standard Model would be the neutrino,
but since neutrinos are relativistic, they cannot explain the
structure formation of the Universe. Therefore, the explanation for
dark matter has to be outside the Standard Model. 

\begin{figure}[h!]
%%\begin{figure}[c][h!]
\begin{center}
\includegraphics[width=0.95\linewidth]{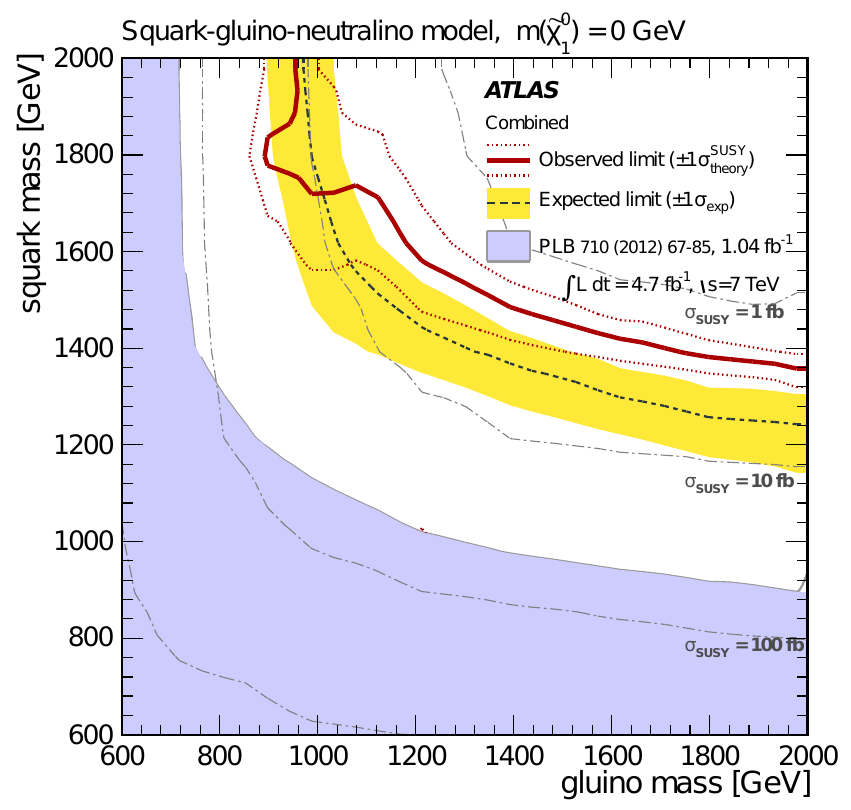}
\caption{Example of constraints to CMSSM from LCH results.}
\label{lhclimits}
\end{center}
\end{figure}

The question of identifying the nature of dark matter has to
be approached from several experimental fronts at the same time. For
instance, there are many searches for supersymmetric partners of the
Starndard Model particles at the Large Hadron Collider. These
experiments have succeded in finding what looks very much like the
Higgs boson, but for the moment, only limits on models beyond the
Standard Model, like Supersymmetry, have been
set (see Figure~\ref{lhclimits} for an example). In the mean time, these results are weakening the arguments in
favor of naturalness~\cite{lhc}. As an example of model to explain the
experimental results, M. Peir\'{o}~\cite{theory2}~proposes an extension of the Next-to-MSSM
(NMSSM) with right-handed neutrinos, in which the the right-handed
component of sneutrinos can be a good DM candidate. The coupling of
these sneutrinos with the Higgs is of the order of the electroweak
scale, making possible the thermal production of these
particles. There is a wide range of predictions for both direct and
indirect detection of DM experiments for very light sneutrinos. The
different predictions are related, generally, with the final state
annihilition in the early universe of these light particles.

%===========================================
%===========================================

\begin{figure}[h!]
\begin{center}
\includegraphics[width=0.95\linewidth]{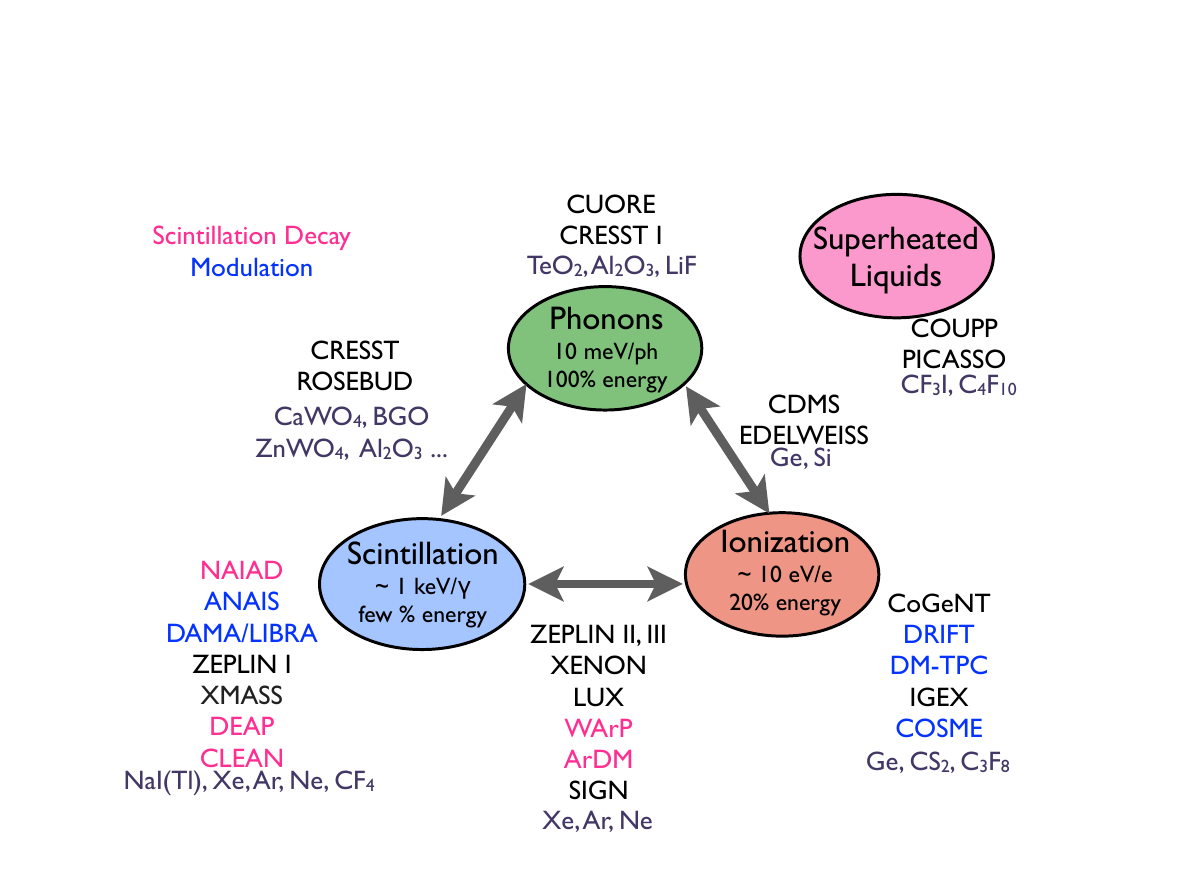}
\caption{Experimental approaches for direct dark matter detection. The
  figure includes both present and future experiments. (Form C. Cuesta
  thesis defense).}
\label{directdetection}
\end{center}
\end{figure}

In any case, a positive signal at LHC
will not be enough to claim the discovery of dark matter, since one of
the main features that a dark matter candidate has to fulfill, the
stability at cosmological scales, cannot be proven in
accelerators. The other two fronts which complement this task are the
so-called direct and indirect searches. In the first case, the
interaction of dark matter particles in the detector are looked
for. The experimental effort here is very broad. Three main signatures
are used, often combined in pairs in order to improve the sensitivity:
scillation (light), ionization (charge) and phonon (heat). Another
technique which is being tested are the superheated liquids
(bubbles). Figure~\ref{directdetection} shows a summary of present and
future experiments, indicating the techniques used in each case. The
present situation is quite puzzling and very interesting. There are
positive results (DAMA/LIBRA, COGENT, CRESST, CDMS-Si) with are in
contradiction, at least for the most simple assumptions, with limits
set by other experiments (XENON100, CDMS-Ge). Moreover, the signal
regions of the positive results are at least in tension among
themselves. In the following subsections we review some of these experiments.

%===========================================
%===========================================

%===========================================
\begin{figure}[h!]
\begin{center}
\includegraphics[width=0.95\linewidth]{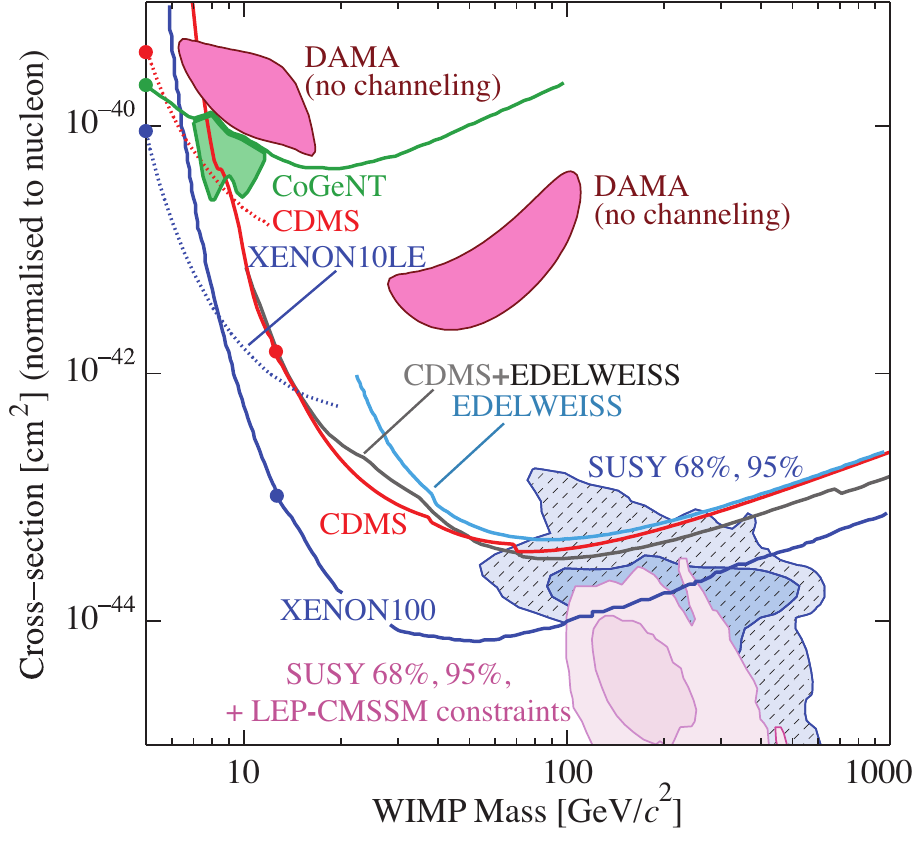}
\caption{Experimental limits and signal favoured regions obtained from
the results of different direct searches of dark matter (source:
Particle Data Group.)}
\label{directlimits}
\end{center}
\end{figure}
%===========================================

Indirect searches look for the particles resulting from the decay or
annihilation of dark matter particles. This includes photons, cosmic
rays and neutrinos. For a summary of the results presented in this
conference for photons and cosmic rays,
see~\cite{reportground}~\cite{reportspace}. As described there, there
are several hints which could be compatible with dark matter. However,
there are also more standard astrophysical scenarios which could
explain these results. Searches for WIMPs in the Sun by neutrino
telescopes would not have this problem if a signal is detected,
since no astrophysical explanation would compete. In this proceeding we
will review the results of the ANTARES neutrino telescope.

\subsection{DAMA/Libra}

The DAMA/LIBRA collaboration~\cite{dama} has deployed about 250~kg of
highly radiopure NaI(Tl) at the Gran Sasso National Laboratory. They
have observed an annual modulation signature which is compatible with
the assumption that it is produced by the asymmetry in the expected
rates of dark matter particle interactions when the Earth is moving
forward or backwards the "wind" of dark matter seen as the Sun moves
around the Galaxy. The requirements for such a signal in this scenario
are that it has to be modulated according to a cosine function, in a
definite low energy range, with a period of one year, whose maximum
should be at June 2nd and to be observed just for single hit events in a
multi-detector set-up. The significance of the observed signal after a
total exposure of 1.17 tons$\times$year is 8.9$\sigma$. The
collaboration has made a lot of effort to discard many possible
systematic effects which could produce a spurious signal, with
negative results: none of the investigated process is able to
simultaneously satisfy all the peculiarities of the signature. The
collaboration plans to continue data taking with a new configuration
which would lower the energy threshold below 2~keV.

%===========================================
\begin{figure}[h!]
\begin{center}
\includegraphics[width=0.95\linewidth]{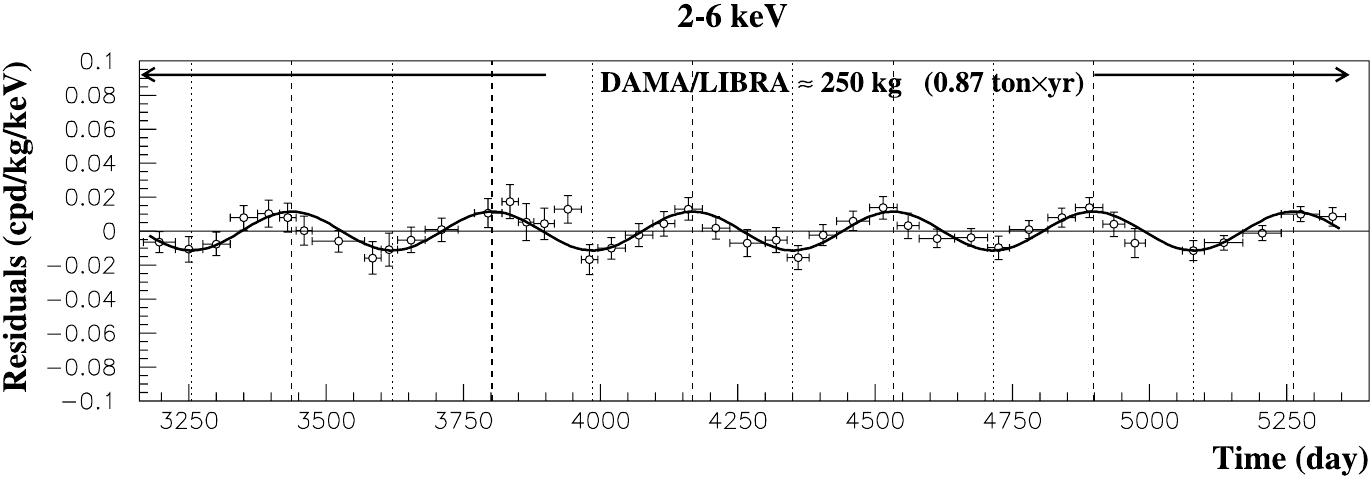}
\caption{Annual modulation (residual rate of single-hit scillation events) observed by the DAMA/LIBRA experiment in
  the 2-6~keV interval.}
\label{damaresult}
\end{center}
\end{figure}
%===========================================

\subsection{ANAIS}

\begin{figure}[h!]
\begin{center}
\includegraphics[width=0.95\linewidth]{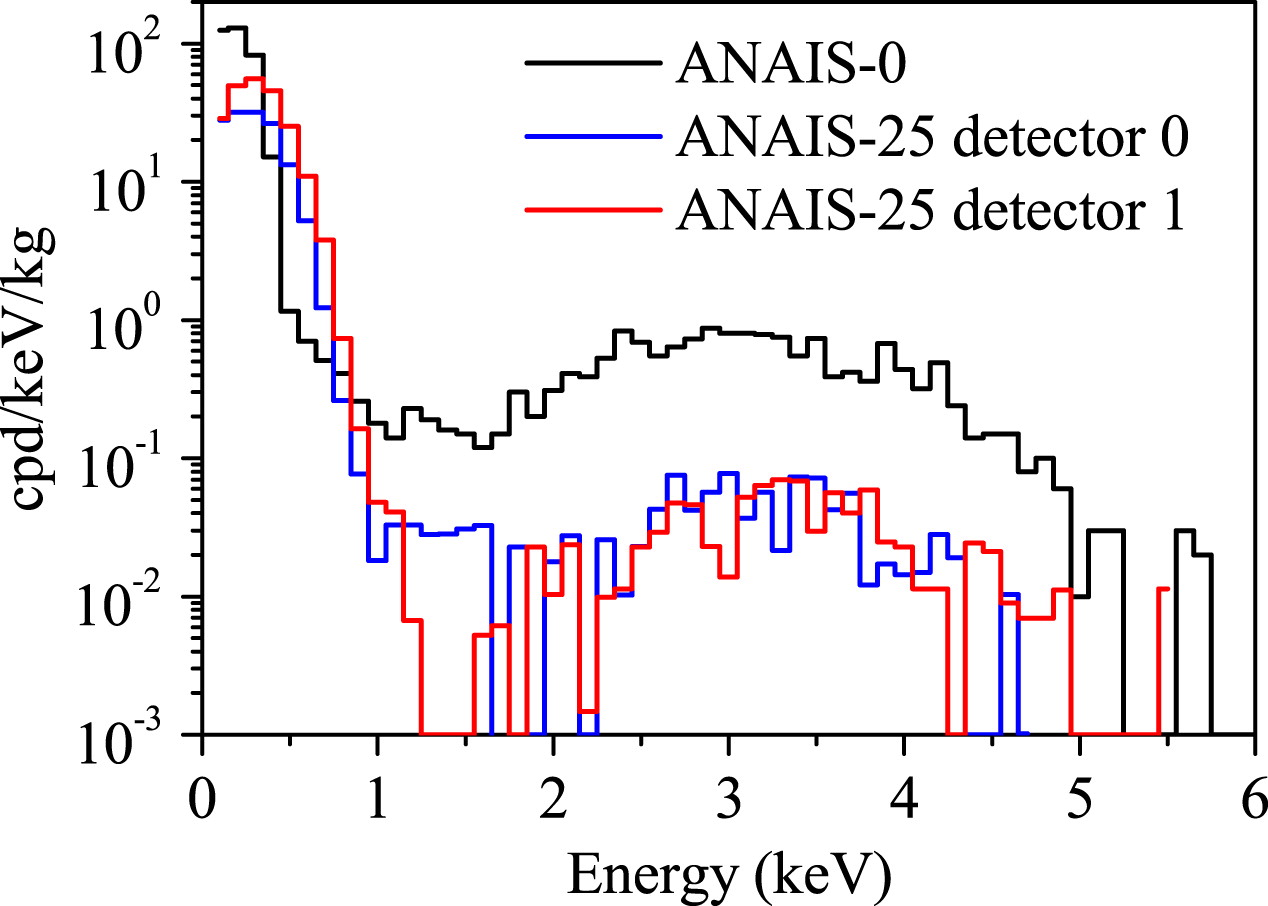}
\caption{Background of potassium-40 as measured for the ANAIS-0 test
  module (9.6~kg) and the ANAIS-D0 and ANAIS-D1 modules (12.5~kg each).}
\label{anaisplot}
\end{center}
\end{figure}

The ANAIS project~\cite{anais} aims to set up a 250~kg
potassium-purified NaI(Tl) detector in the Canfranc Undergournd
Laboratory. The goal is to confirm the annual modulation observed by
the DAMA/LIBRA experiment. The target mass will be divided in 20
modules containing of a 12.5~kg cristal coupled to two
photomultipliers. A first test module (ANAIS-0, 9.6~kg) was
characterized at the Canfranc Laboratory. These measurements allowed,
among other things, to optimize the event selection strategy, to
design the calibration method and to test the acquisition code and
electronics. Also several photomultiplier models were tested. From
December 2012, measurements on two 12.5~kg modules (ANAIS-D0 and
ANAIS-D1) have been performed to determine the bulk contamination. The
potassium-40 bulk content has been measured as 41.7$\pm3.7$~ppb, by
performing a search for coincidences of 3.2~keV in one detector and
1460.9~keV in the other (see Figure~\ref{anaisplot}. Concerning the $^{232}$Th and $^{238}$U
chains, a high rate (3.15 mBq/kg) have been observed, attribuable to
$^{238}$U chain with broken equilibrium. The contribution from
cosmogenic activated isotopes can also be clearly identified and its
decay along the first months of data taking is observed. A background
model is under development to reproduce the observed rates.

%===========================================
%===========================================

\subsection{DarkSide}

\begin{figure}[h!]
\begin{center}
\includegraphics[width=0.95\linewidth]{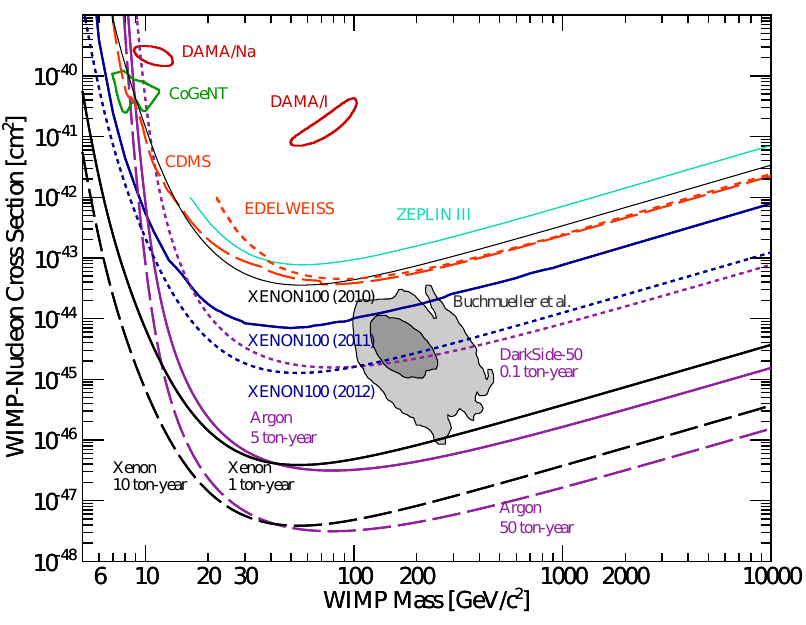}
\caption{Expected sensitivity for DarkSide-50 to the WIMP-nucleon scattering
  cross section, compared to other experiments.}
\label{darksideplot}
\end{center}
\end{figure}

The DarkSide-50 detector~\cite{darkside} is a two phase Argon TPC intalled in the the
Gran Sasso Laboratory. The experimental design follows a null
background strategy, so that an unambiguous discovery could be claimed
with just a few WIMP candidates. To this end, several strategies are
implemented, apart from setting the detector deep underground. First,
to reject electro/gamma background by using the information on the
shape of the scillation signal, the ratio between ionization and
scintillation and the 3D location of the event. Second, to use
underground argon depleted in $^{39}$Ar, which is 150 times
more radiopure than atmospheric argon. Third, to use a liquid
scintillator as active neutron veto. Fourth, to use a water tank for
detection of muons and shielding against external neutrons. The total
expected background with this setup is less than 0.1 per ton and
year. As a first step to estimate the background and validate the
setup, a smaller version has been built (DarkSide-10, 10~kg target), which has been
successfully operated for one year. The DarkSide-50 detector (50~kg
target) is under comissioning and the expected sensitivity is
indicated in Figure~\ref{darksideplot}, compared with other
experimental results.

%===========================================
%===========================================

\subsection{ANTARES}

\begin{figure}[h!]
\begin{center}
\includegraphics[width=0.95\linewidth]{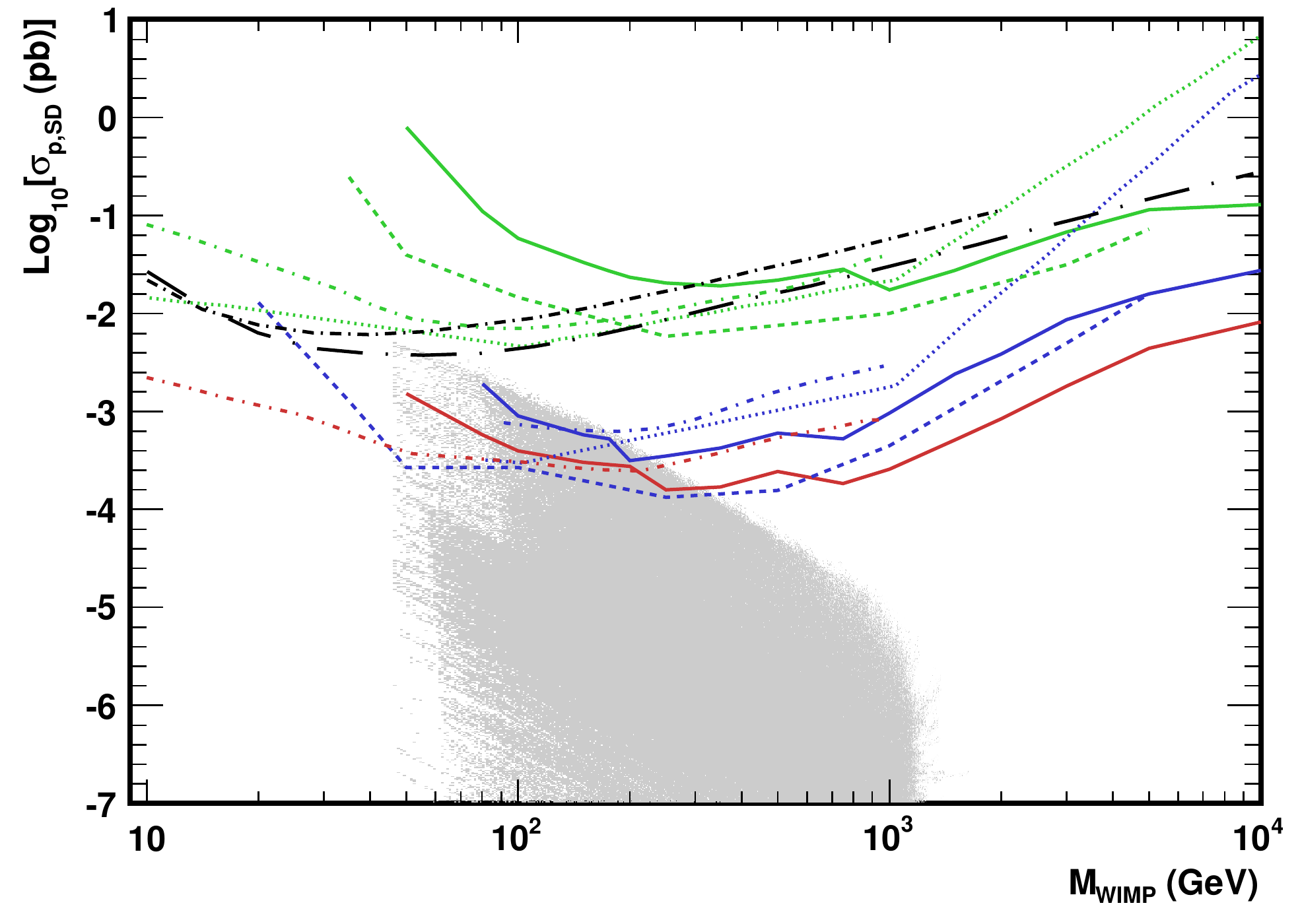}
\caption{$90$\% CL upper limits on the SD WIMP-proton cross-sections
  (plots on the left and right, respectively) as a function of the
  WIMP mass, for the three self-annihilation channels: $b\bar{b}$
  (green), $W^{+}W^{-}$ (blue) and $\tau^{+}\tau^{-}$ (red), for
  ANTARES 2007-2008 (solid line) compared to the results of other
  indirect search experiments: Baksan $1978-2009$ (dash-dotted lines),
  Super-Kamiokande $1996-2008$ (dotted lines) and IceCube-$79$
  $2010-2011$ (dashed lines) and the result of the most stringent
  direct search experiments (black): SIMPLE $2004-2011$ (short
  dot-dashed line in upper plot), COUPP $2010-2011$ (long dot-dashed
  line in upper plot) and XENON100 $2011-2012$ (dashed line in lower
  plot). The results of a grid scan of CMSSM-7 is included (grey
  shaded area) for the sake of comparison.}
\label{antaresplot}
\end{center}
\end{figure}

The ANTARES neutrino telescope is a three-dimensional array of
885 photomultipliers installed in the Mediterranean sea (more details
in~\ref{sec:antares}). In addition to the detection of neutrinos of
astrophysical origin, the indirect detection of dark matter is a major
scientific goal of the experiment. The idea is that WIMPs would
accumulate in massive objects like the Sun and their annihilation
would produce (directly or indirectly) high energy neutrinos. One of
the advantages of such kind of detectors is that a potential signal
from the Sun would be a very clean indication of dark matter, since no
other astrophysical explanations are envisaged, contrary to the
situation of other indirect searches (like positrons or gammas) for
which the rejection of astrophysical sources is much more difficult
(i.e. pulsars in the case of positrons). ANTARES has already produced
limits for neutralinos in the Sun with data taken during 2007-08~\cite{antares2} (a
new analysis with 2007-2012 data is also about to be completed, as
well another one looking for a signal in the Galactic
Center). The analysis is a binned search in which the search cone has
been optimized for optimum sensitivity. The signal is simulated with
the WimpSim package and the background is directly obtained by
scrambling data, which has the advantage of lower systematic
uncertainties. There are two sources of background. On the one hand,
the muons produced by cosmic rays in the atmosphere, which can be
greatly rejected by selecting only upgoing events. Still, an important
number of atmospheric muons which are misreconstructed as upgoing
remain, so further cuts on the quality of the reconstructed track are
needed. A second source of background are the atmospheric neutrinos,
also produced by cosmic rays in the atmosphere. This is an irreducible
background, but distributed smoothly in the sky, contrary to a
potential signal, which should concentrate in the Sun's
direction. Figure~\ref{antaresplot} shows the limits on the spin
dependent WIMP-proton cross section. For this case, the limits are
better than those obtained by direct search experiments. 

%===========================================
%===========================================

\section{Underground experiments}
\label{underground}

The background produced by cosmic rays (in particular atmospheric
atmospheric muons since they are the most penetrating component) makes
impossible to carry out experiments of low signal on the surface. This
is way it is important to shield the detectors by installing them deep
underground. In order to reduce costs, the location is typically in
already made excavated sites, like old mines or tunnels through
mountains. This is the case of some of the experiments looking for
dark matter presented in the previous section and also the ones
presented in this section (those non related with dark matter), all of
them installed in the Gran Sasso Laboratory.

\subsection{Borexino}

The Borexino experiment~\cite{borexino} is a low background neutrino
detector located at Gran Sasso Laboratory whose main goal is the
detection of sub-MeV solar neutrinos. The signature for this search is
neutrino electron scattering produced in a nylon vessel containing
278 tons of organic liquid scintillator. This vessel is sourrounded itself by a water tank
for gamma and neutron shielding. One of the most relevant and recent
results from Borexino is the first detection of $pep$ solar neutrinos. This
measurement is important because it offers a more stringent test on
oscillation models. Another recent
result is the strongest constraint on the CNO solar neutrino flux,
which is key in order to distinguish between high and low metallicity
models. In order to make possible these analyses and given the
involved fluxes (ten times lower than those for $^{7}$Be,
it is crucial to reject the cosmogenic $^{11}$C,
which is the dominant background in the 1-2~MeV range. This is done by
a threefold coincidence technique and pulse shape discrimination. In
addition to solar neutrinos, the scientific program of Borexino also
includes geoneutrinos (antineutrinos emitted in beta decays of
naturally occuring radioactive isotopes in the Earth's crust and
mantle). The location of Borexino offers the advantage of a moderate
contamination from nuclear reactors. Values for the fluxes from
Uranium and Thorium have been obtained.

%===========================================
\begin{figure}[h!]
\begin{center}
\includegraphics[width=0.95\linewidth]{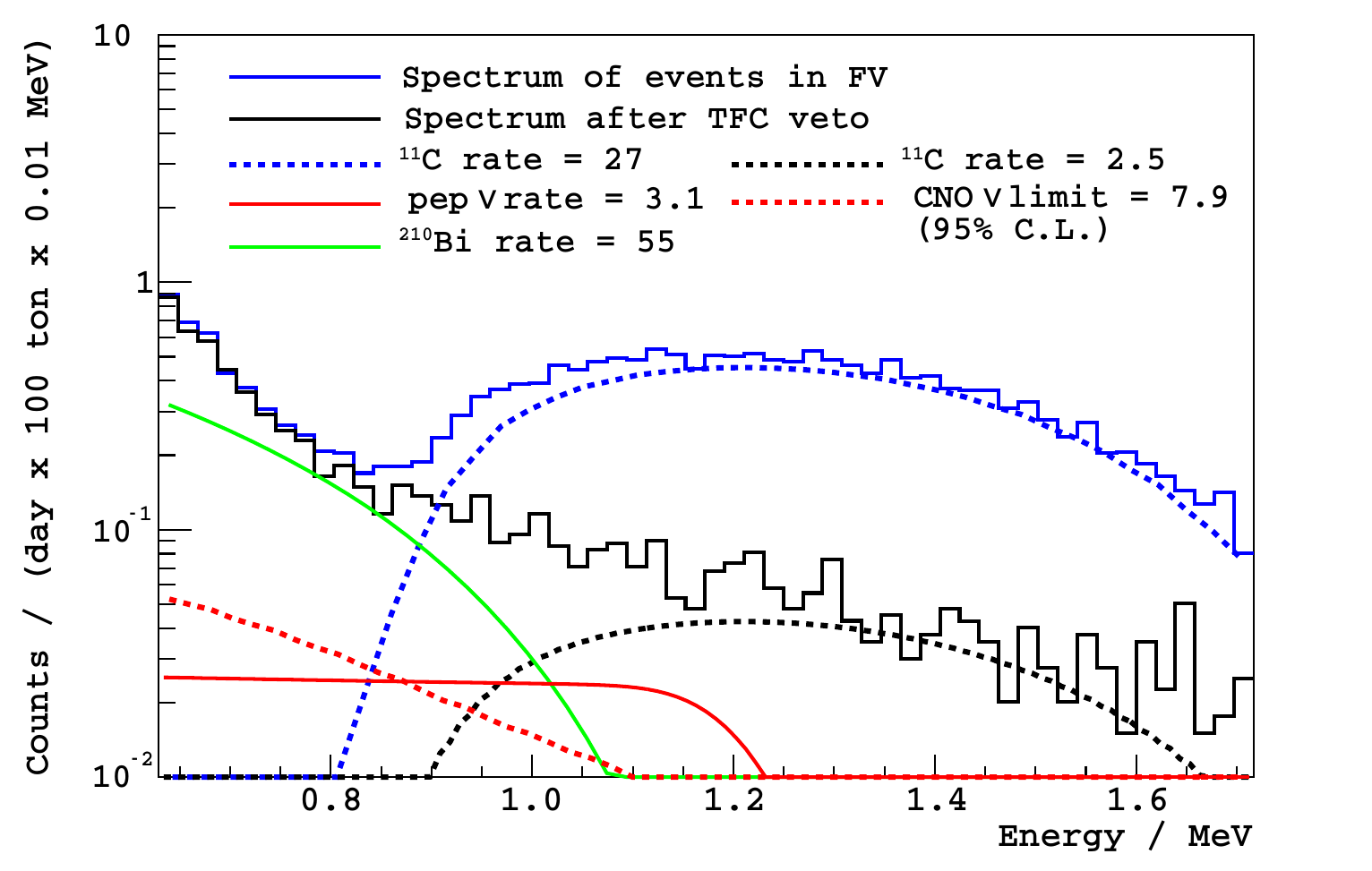}
\includegraphics[width=0.95\linewidth]{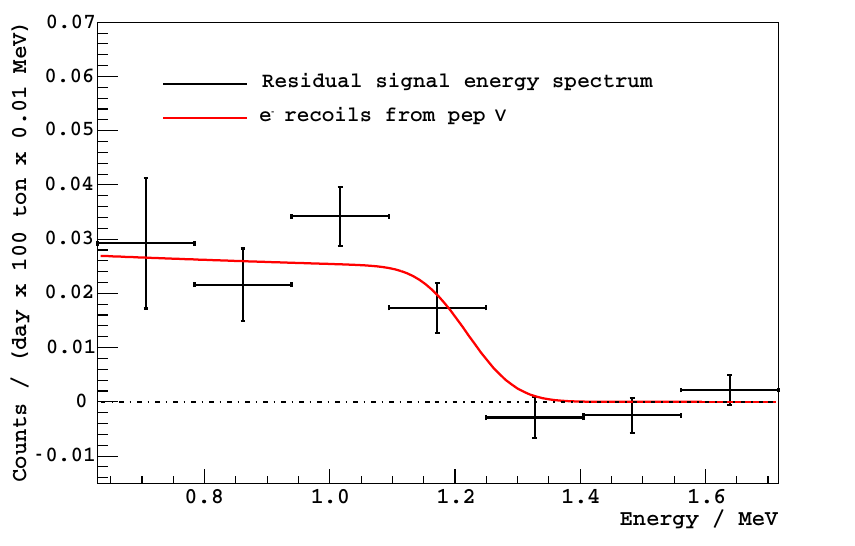}
\caption{Top: Energy spectrum obtained in the Borexino detector for the search
of e$^-$ recoils from pep neutrinos. The plot shows both the spectrum
before and after applying a triple coincidence cut used for reducing
the $^{11}$C background. Botton: Residual spectrum after substracting
the background.}
\label{borexinoplot}
\end{center}
\end{figure}
%===========================================

\subsection{GERDA}

\begin{figure}[h!]
\begin{center}
\includegraphics[width=0.95\linewidth]{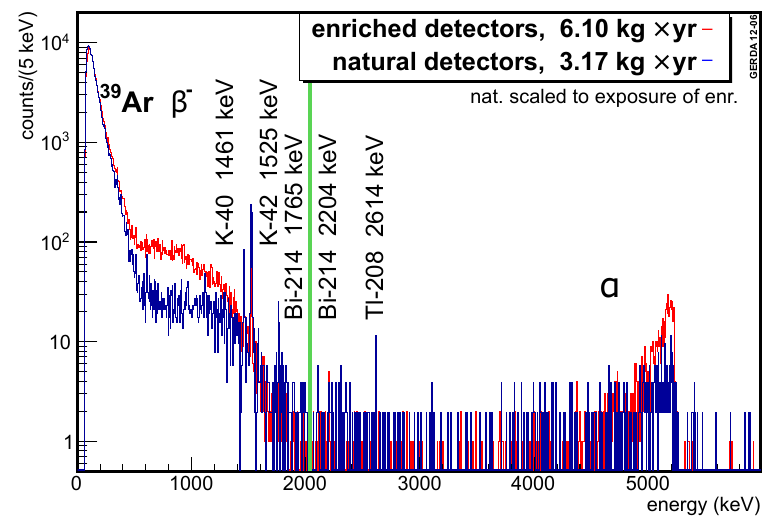}
\caption{Spectrum of enriched (red) and natural (blue) diodes for the
  GERDA experiment. The region $\pm20$~keV around
Q$\beta\beta$ is blinded, as indicated in green in the figure.}
\label{gerdaplot}
\end{center}
\end{figure}

The main goal of the GERDA experiment~\cite{gerda}, located in the
Gran Sasso Laboratory, is the detection of
neutrinoless double beta decay in $^{76}$Ge, which would prove the
Majorana nature of neutrinos. The detector uses isotopically enriched
Ge diode embeded directly in liquid Argon, which serves for cooling,
radiation shielding and active veto. The Phase I of the experiment has
used $\sim$18~kg of enriched Germanium, taking data in the period
November 2012 - May 2013, with a total exposure of 20.9~kg~year for the
enriched detectors. For the Phase II, to start in 2013, 30 additional
enriched BEGe (Broad Energy Germanium) detectors will be added, for a total of 48~kg of enriched
Germanium. For this phase, the goal is to have background at the level
of a few counts/(keV kg yr), with a total exposure of $>$100~kg
year and a liquid Argon scintillation veto. The detector includes a muon Cherenkov veto, a detector
anti-coincidence veto, a pulse shape discrimination. In Figure~\ref{gerdaplot} the energy spectrum
after selection cuts is shown. Several background components can be
identified, like $^{39}$Ar, $^{42}$K, gamma lines and the surface alphas. The
collaboration has followed a blinding strategy for the data, in order
to avoid selection biases. The region in $\pm20$~keV around
Q$_{\beta\beta}$ will be soon unblinded.

%===========================================
%===========================================

\subsection{LUNA}

The 400~kV LUNA accelerator~\cite{luna} is located in the Gran Sasso Laboratory and
provides beams of protons, $^{4}$He and $3$He. This accelerator
allows to measure nuclear cross-sections which are important
to understand several astrophysical phenomena like nucleosynthesis,
energy production in stars or stellar evolution models. Direct measurements of
cross-sections are important in order to avoid problems from
extrapolations from higher energies due to resonances. One example of
recent result obtained with LUNA is the measurement of resonance
strengths $\omega\gamma$ in the reaction 25 Mg(p, $\gamma$ )26 Al. The
obtained results imply a reduction of the estimated contribution of Wolf-Rayet stars to the galactic production of $^{26}$Al.

%===========================================
\begin{figure*}[h!]
\begin{center}
\includegraphics[width=0.95\linewidth]{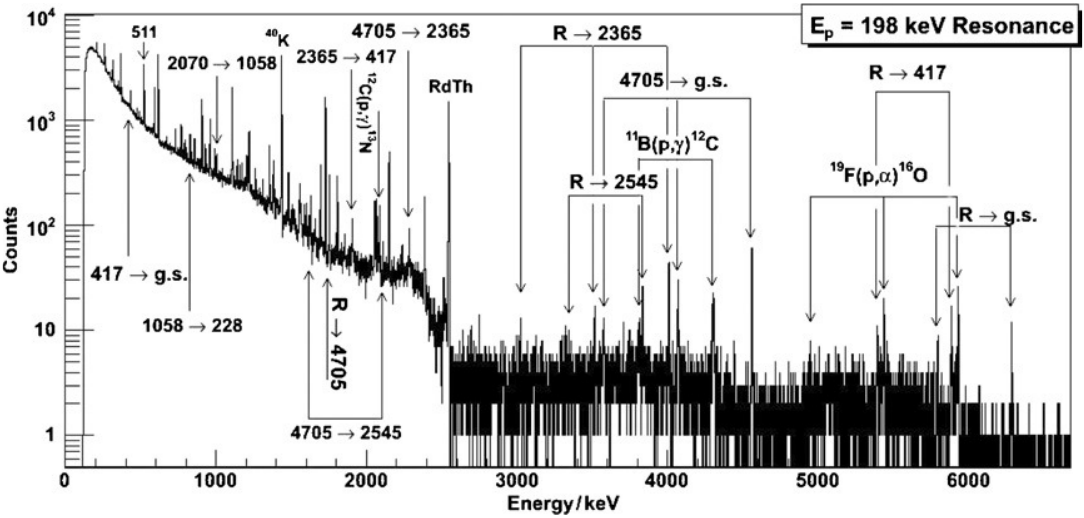}
\caption{$\gamma$-ray spectrum taken at the E = 189 keV 25 Mg(p, $\gamma$ )26 Al resonance showing the most prominent primary transitions, important secondary transitions and some $\gamma$-ray background lines.}
\label{lunaplot}
\end{center}
\end{figure*}
%===========================================

\section{Neutrino telescopes}
\label{neutrino}

Neutrino astronomy has become an important tool both for astrophysics
and particle physics. In order to study de Universe at high energies,
traditional probes like photons and protons have important
drawbacks. High energy photons interact with matter and radiation,
which limits their range. Protons can be also absorbed by radiation
and matter. Moreover, protons, being charged particles, are deflected
by magnetic fields, which erases the directional information. This is
why, more than one hundred years after the discovery of the
extraterrestrial origin of cosmic rays, their sources are still
unkown, in particular at high energies. Neutrinos, on the other hand,
are neutral and only interact weakly. The price to pay is that large
detector volumes are needed given the weak cross section and fluxes
involved. In any case, gamma astronomy, cosmic rays and neutrinos are
closely related. In processes where cosmic rays are expected to be
produced (via interaction of those with matter or radiation, yielding
pions) gamma rays and neutrinos are also expected, from the decay of
the neutral and charged pions respectively. Therefore, the main goals
of neutrino telescopes is the understanding the origin of cosmic rays
(since neutrinos are also produced and they point
back to the sources) and disentangle between the hadronic (where
neutrinos are also produced) or leptonic (no neutrino production)
mechanisms which can explain the gamma rays observed from several
astrophysical sources. Another important motivation is to look for
dark matter from sources like the Sun, as explained before. There are
two main signatures in neutrino telescopes. First the long tracks observed
by the array of photomultipliers of the
Cherenkov photons induced by the relativistic muons produced in the CC
interactions of neutrinos around/inside the detector. The second
signature correspond to CC interactions of electron and tau
neutrinos as well as the NC interactions of any flavour, where a
shower is produced and observed as a bright sphere of light. Given the
large volumes of transparent media required for such detectors, two
options arise: the Antarctic ice or lakes/sea water. For very high
energies, i.e. above tens of PeV a different strategy has to be
followed, since the fluxes are too low and require larger detection
volumes. To instrument such large volumes with photodectectors would
be too expensive, given the attenuation length of light. Therefore,
accoustic or radio signals have to be used. An analysis on how to
discriminate between diferent neutrino flavors in radio neutrino
experiments can be found here~\cite{flavors}.

In the following, we review the status and some results of the water
Cherenkov neutrino telescopes in the world. 

\subsection{IceCube}

\begin{figure}[h!]
\begin{center}
\includegraphics[width=0.95\linewidth]{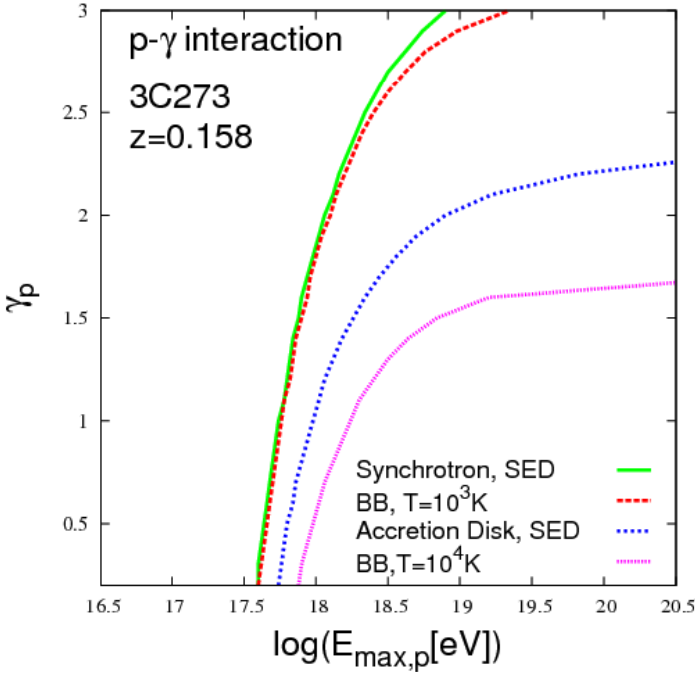}
\caption{Constraints on (E$^{max}$,p,$\gamma$p) deduced in the
  p$\gamma$ model for proton interactions with a soft photon
  distribution corresponding to the acretion disk spectrum~\cite{icecube1}.}
\label{icecubeplot1}
\end{center}
\end{figure}

The IceCube detector is installed in deep ice in the Antarctic. It
consists of 5160 photomultipliers distributed along 86 lines deployed
at a depth between 2.5 and 1.5~km under the surface. There is also a
complementary array of water Cherenkov tanks on the surface, IceTop, with a
total of 324 photomultipliers. The detector was completed in 2011,
although data taking was possible with only part of the detector
deployed, producing a rich set of physics results. For instance, the
observations when only 40 lines of IceCube were operative (IC40)
produced a set of limits in the neutrino flux for a list of blazars
that had been observed by Fermi. Assuming that the flux of photons are
produced by hadronic mechanisms, limits on the primary proton flux can
be set~\cite{icecube1}. An example of the limits obatined for 3C273,
for the case of p-$\gamma$ interaction and for the acretion disk
spectrum model is shown in Figure~\ref{icecubeplot1}.

%===========================================
%===========================================

\begin{figure}[h!]
\begin{center}
\includegraphics[width=0.95\linewidth]{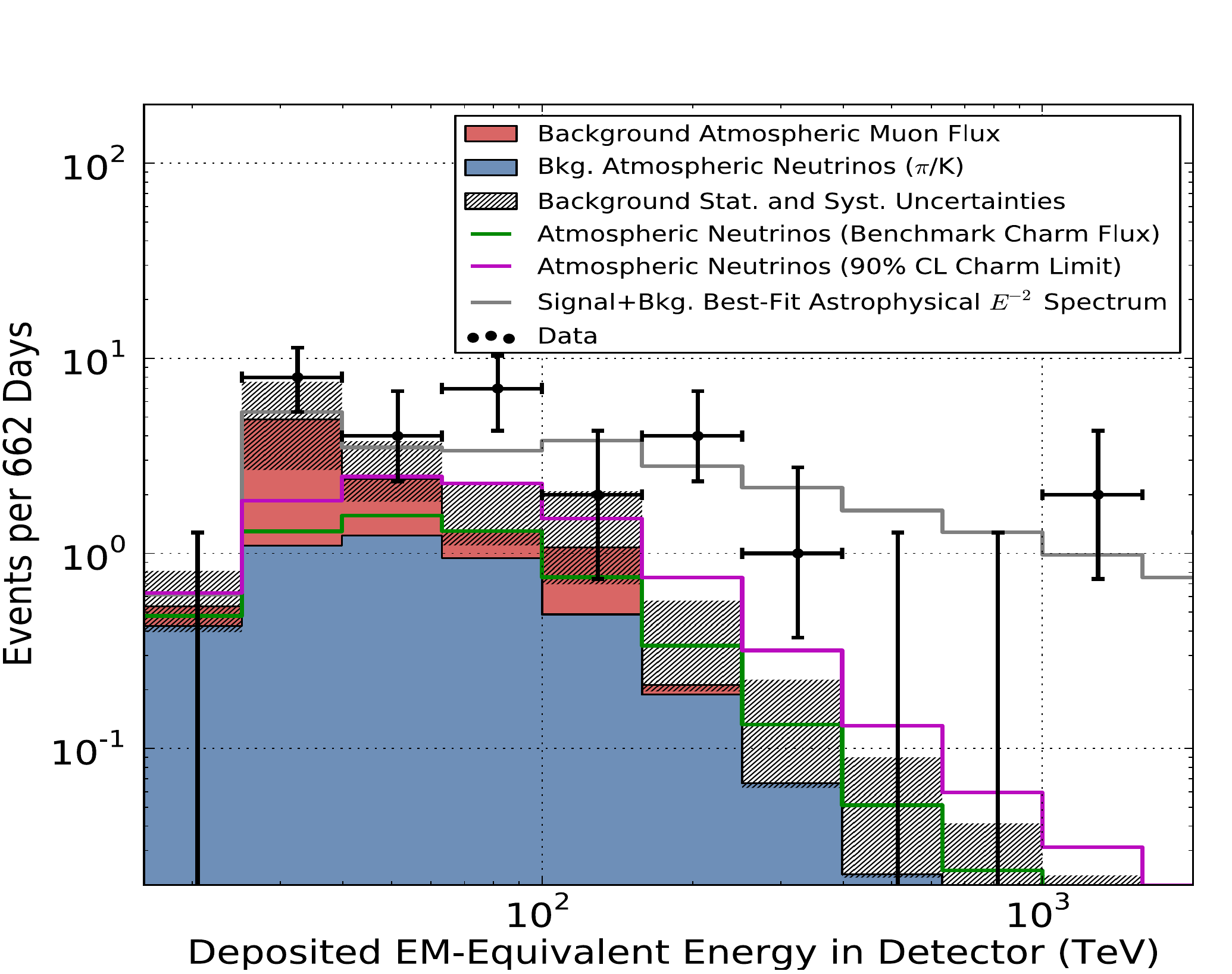}
\caption{Distribution of desposited energy in the IceCube detector for
the HESE analysis.}
\label{icecubeplot2}
\end{center}
\end{figure}

The data of 2011-2012, when the IceCube detector was almost finished
(IC79) or already completed (IC86), has produced a result which seems to be the
first evidence of extraterrestrial neutrinos~\cite{icecube2}. The first hint came from
the observation of cascade events with reconstructed energies just
above 1 PeV. These two events were found when looking for cosmogenic
neutrinos, but their energies are too low to compatible with such
origin. On the other hand, this suggested to lower the energy
threshold to look for more events in the 100~TeV - 1~PeV region. The
event selection strategy was based on looking for High-Energy Starting
Events (HESE), i.e. events for which the interaction vertex is
contained inside the detector. In order to make such a selection, the
external lines of the detector are used for vetoing, ensuring that the
depostion of energy starts inside the detector. The adavantage of this
strategy is that the background is greatly reduced both down-going
atmospheric muons and atmospheric neutrinos (which are usually
acompained by muons produced in the cosmic ray showers producing
them). This allows the acceptance of the detector to cover
homogenously the whole sky. The energy threshold, contrary to previous
analysis looking for donward events, where it was $\sim$PeV, it is
about 80~TeV. The expected background for the period of the analysis
(May 2010 - May 2012) is 6.0$\pm$3.4 atmopheric muons and
6.1$^{+3.4}_{-3.9}$ atmospheric neutrinos (including conventional and
prompt contributions). The total background is therefore 12$^{+4.5}_
{-3.9}$. The number of observed events is 28, including the two PeV
events already observed in the previous analysis (see Figure~\ref{icecubeplot2}). The total
signficance (without including the two previous PeV events) is
4.3$\sigma$. In addition to the rate, there are other hints which
support the extraterrestrial origin of this excess. Out of the 28
events, 21 of them are cascades and 7 are tracks. This is compatible
with the cosmic origin (when the effect of oscillation is taken into
account) but not with the conventional atmospheric neutrino
expectations. The spectrum is consistent with a E$^{-2}$ index (with a
cutoff about 1.6 PeV).  This is a factor two below the previous limits
of IceCube for the standard diffuse flux analysis. Moreover, the
vertices are homogenously distributed within the detector,
disfavouring a possible leak of atmospheric origin.

%===========================================
%===========================================

\begin{figure}[h!]
\begin{center}
\includegraphics[width=0.95\linewidth]{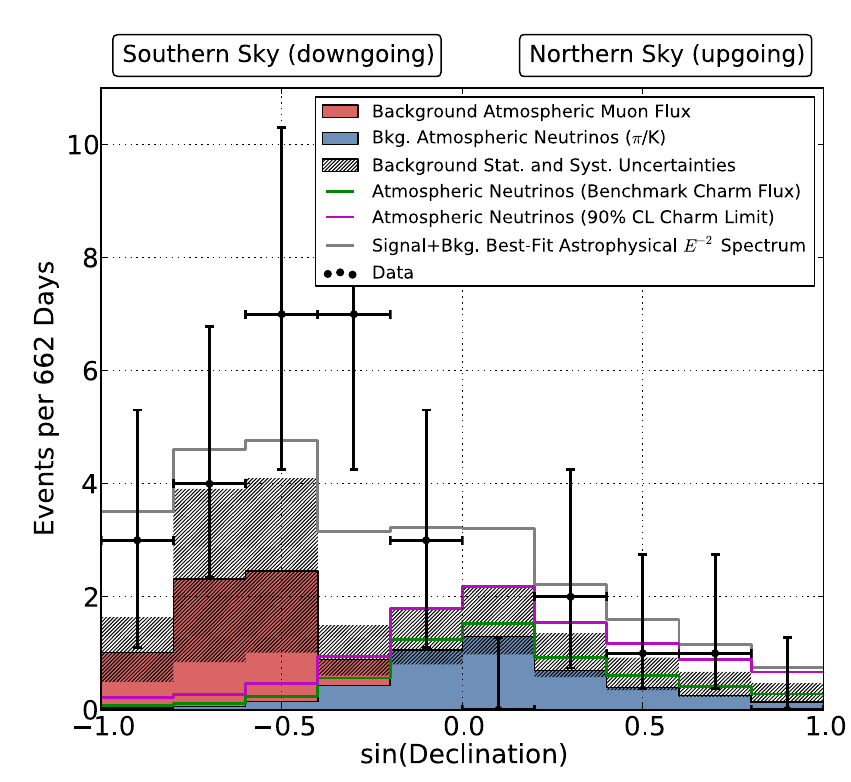}
\caption{Declination distribution of the 28 events observed by IceCube
  in the HESE analysis.}
\label{icecubeplot3}
\end{center}
\end{figure}

Concerning the
distribution of the events in the sky, there is an excess from the
Southern hemisphere~\ref{icecubeplot3}. Part of this excess is expected due to the
absorption on Earth for upgoing (Northern origin) events. There is
still an intriguing excess when the previous effect is taken into
account, although not very significant, so the mesured declinations
seem to be compatible with a diffuse distribution.

%===========================================
%===========================================

\subsection{ANTARES}
\label{sec:antares}

\begin{figure}[h!]
\begin{center}
\includegraphics[width=0.95\linewidth]{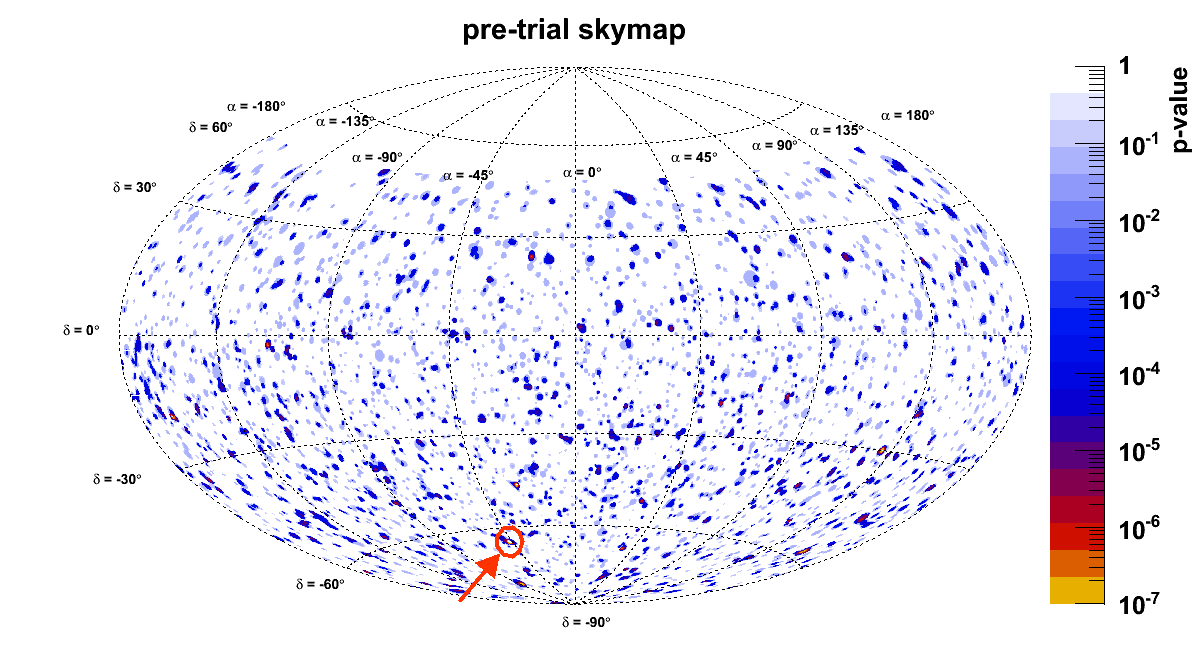}
\caption{Skymap in equatorial coordinates of the pre-trial p-values
  observed by ANTARES.}
\label{antaresplot1}
\end{center}
\end{figure}

The ANTARES detector~\cite{antares3} is located 2500~m deep in the Mediterranean Sea,
at 40~km of the French city of Toulon. This location offers a great
visibility of the Galactic Center and most of the Galactic Plane. It consists of 885
photomultipliers distributed along 12 vertical anchored at the sea bed
and kept vertical by a buoy. The detector was completed in 2008. The
scientific output has been very rich, including search for point
sources, cosmic diffuse fluxes, first oscillation effects observed by
a neutrino telescopes, search for correlations with transient sources
like gamma ray bursts, blazars and microquasars and with gravitational
waves and also searches for more exotic phyiscs like monopoles or
nuclearites. In the following we review a few of these
results. Figure~\ref{antaresplot1} shows the result of the all-sky
search for point-like neutrino sources, using 2007-2010 data (813
active days). The most significant cluster is
indicated by the circle (centered at $\delta$=-46.5$^{\circ}$ and
R.A.=-65.0$^{\circ}$. The post-trial significance is 2.2$\sigma$. In
paralel to this all-sky analysis, a search in specific locations of
good neutrino candidates sources has been made, without any relevant
excess.

%===========================================
%===========================================

\begin{figure}[h!]
\begin{center}
\includegraphics[width=0.95\linewidth]{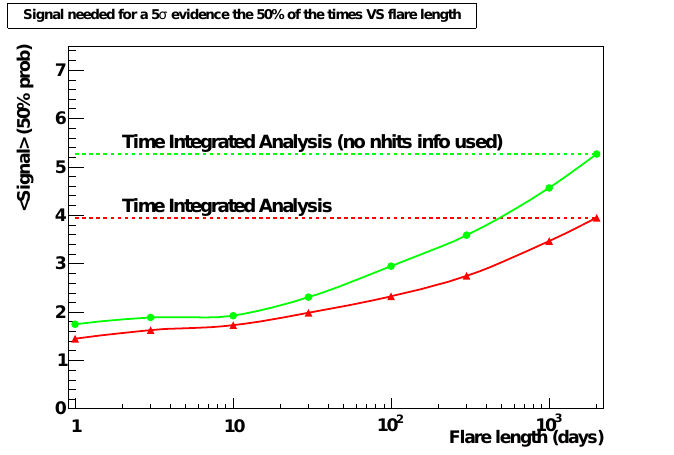}
\caption{Number of events needed for a 5$\sigma$ discovery (for a
  source at $\delta$=-40$^{\circ}$) when the time information is used
  for transient sources (solid line) compared to the case of not using
  the timing information (dashed line), computed using (red) or not
  (green) the energy information in the likelihood.}
\label{antaresplot2}
\end{center}
\end{figure}

For sources which are transient~\cite{antares1} (both single-event cases like
gamma-ray bursts and flaring sources like AGNs or micro-quasars), the
information about when the neutrinos are expected allows to reduce
greatly the background and improve the sensitivity, as shown in
Figure~\ref{antaresplot2}. Several searches have been carried out in
ANTARES following this strategy, including 287 gamma-ray bursts
occuring during 2008-2011, 10 flaring blazars from the 2008 Fermi
catalogue and 8  micro-quasars flaring during 2007-2010. A coincidence
was observed (0.56$^{\circ}$) with a flare of the blazar 3C279, which
gives a post trial p-value of 0.1. No other coincidences has been observed.

%===========================================
%===========================================

\begin{figure}[h!]
\begin{center}
\includegraphics[width=0.95\linewidth]{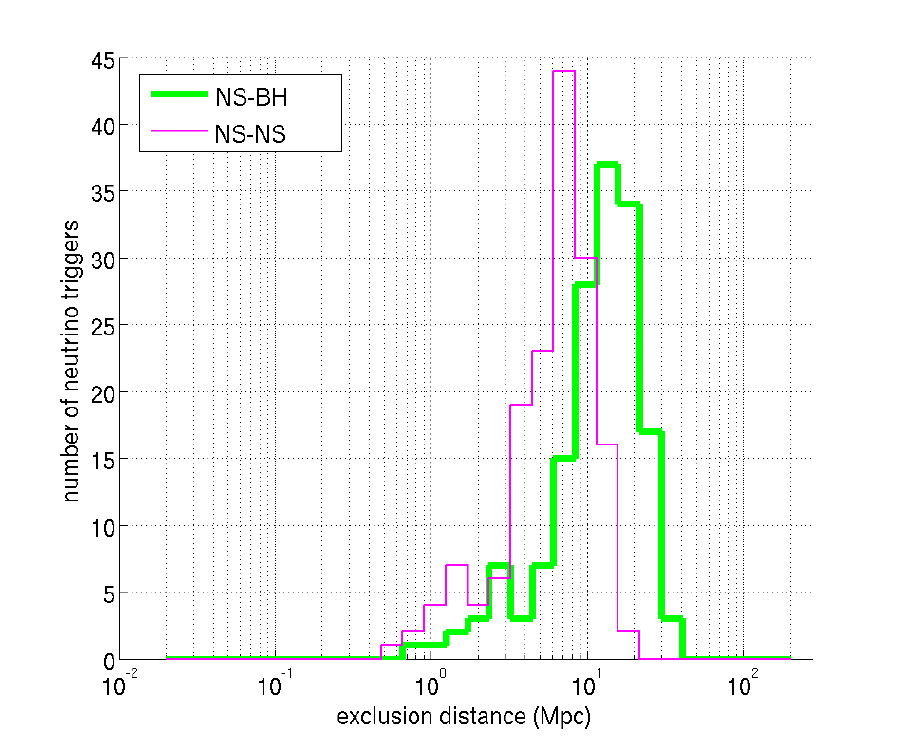}
\caption{Distance exclusions for two potential scenarios for
  gravitational waves and high energy neutrinos joint production:
  Neutron Star - Black Hole (NS-BH) and Neutron Star - Neutron Star (NS-NS).}
\label{antaresplot3}
\end{center}
\end{figure}

Another interesting analysis using the idea of correlations to increase
sensitivity concerns the search for gravitational waves in coincidence
with high energy neutrino events~\cite{gwhen}. 216 neutrino
triggers in the data of 2007 have been analized with gravitational wave data of the LIGO
and VIRGO experiments. One event had a false alarm probability of
0.004, which is perfectly compatible with background when the number
of trials is taken into account. Assuming some assumptions in the
gravitational wave generation, exclusion distances can be set, as
shown in Figure~\ref{antaresplot3}.

%===========================================
%===========================================

\subsection{KM3NeT}

The KM3NeT project~\cite{km3net} is the joint effort of the ANTARES, NEMO and NESTOR
collaborations to build a multi-km$^3$ detector in the Mediterranean
Sea. There has been an extensive program of R\&D for the design of the
detector and the associated technologies (deployment of lines, etc.)
One of the novelties fruit from this R\&D program is the multi-PMT
optical module: instead of a large photochathode area PMT, the optical
modules will contain 31 3'' PMTs, which offers an increased total
photocathode area, a better 1-vs-2 photoelectron separation and
directionality information.

\begin{figure}[h!]
\begin{center}
\includegraphics[width=0.95\linewidth]{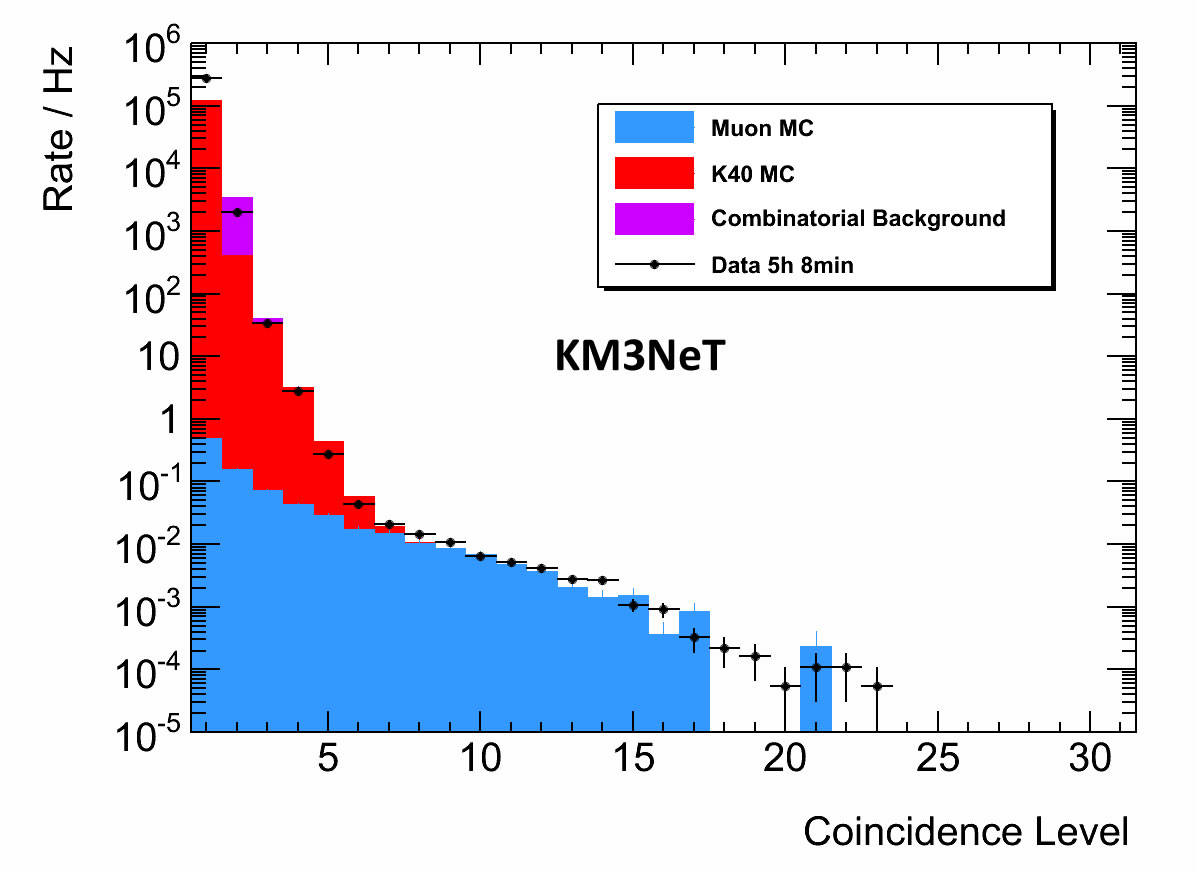}
\caption{Coincidence rates for the KM3NeT PPM-DOM installed in the
  Intrumentation Line of ANTARES.}
\label{km3plot}
\end{center}
\end{figure}

Among the recent progess made by the collaboration for testing the
elements of the detector, we can mention the mounting of one multi-PMT
optical module in the so-called Instrumentation Line of ANTARES,
installed in April 2013 at the ANTARES site. The multi-PMT optical
module is fully operational and working correctly (see Figure~\ref{km3plot}). Moreover, a test
for the deplyment and unfurling of a prototype KM3NeT line was
successfully carried out in April 2013.

%===========================================
%===========================================

\subsection{NEMO}

The NEMO collaboration~\cite{nemo} has carried a R\&D program to
develop a cubic kilometer detector close to Capo Passero
(Sicily). In the so-called NEMO Phase II Project, the Capo Passero
site has been evaluated and a tower prototype has been successfully
installed in March 2013. This
tower contains eight bars (8~m long)  with a vertical separation of
40~m and 4 PMTs in each bar. Moreover, it contains several calibration
instruments, including hydrophones for positioning and a laser beacon
and a LED beacon, used for time calibration. The data taking is
going-on smoothly. The KM3Net-Italy project is developing the
communication architecture and new electronics for KM3NeT.

%===========================================
\begin{figure}[h!]
\begin{center}
\includegraphics[width=0.60\linewidth]{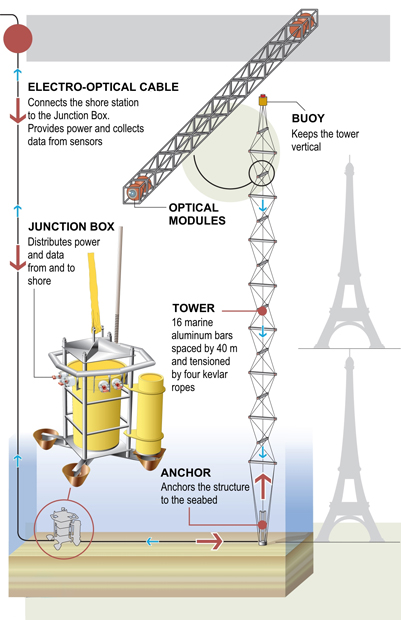}
\caption{Scheme of the Nemo Phase II tower, installed in Capo Passero
  in March 2013.}
\label{nemoplot}
\end{center}
\end{figure}
%===========================================

\subsection{Baikal-GVD}

The GVD project~\cite{baikal} is the extension of the Baikal detector, installed in
the Baikal lake in Siberia. The Baikal neutrino telescope pioneered
neutrino astromony. One of the advantages of the site is the freezing
of the lake surface during winter, which allows to install and test
equipment and to make all the connections on dry. The GVD extension
would consist of $\sim$10,000 additional PMTs instaled in 216
strings. The active part of the detector would be installed at depts
between 600 and 1300~m. The total instrumented volume would be
1.5~km$^{3}$. Several prototype lines have already been installed and
tested. In 2012, a cluster with 3 full-scale lines (with 24 PMTs) was
installed and taking data since April 2012. In 2013, 3 full-scale
strings (72 PMTs), with updated electronics, were installed and are
taking data since April 2013. The next goal is to install by 2015
eight full scale lines.

%===========================================
\begin{figure}[h!]
\begin{center}
\includegraphics[width=0.95\linewidth]{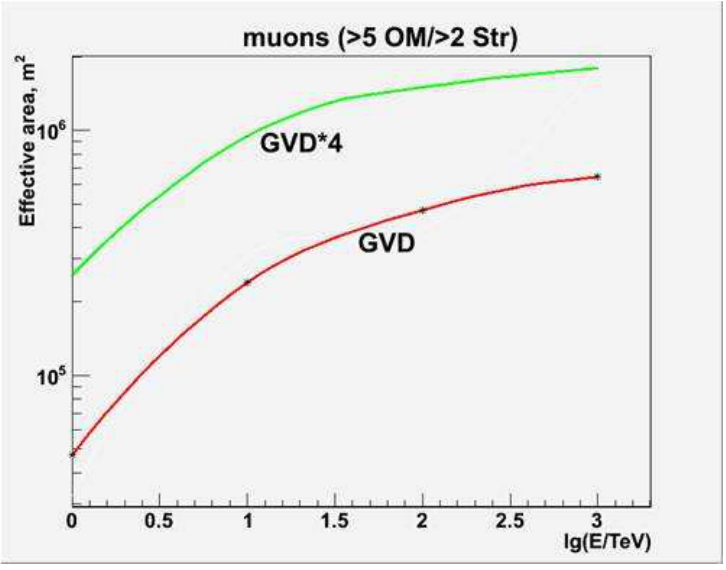}
\caption{Projected effective area as a function of the energy for the
  Baikal-GVD detector at two different states. The total planned
  number of PMTs is 10368.}
\label{baikalplot}
\end{center}
\end{figure}
%===========================================

\section{Conclusions}
\label{conclusions}

There are many frontier physics topics for which it is necessary to be
protected from the atmospheric muon cosmic ray background and
therefore to install detectors deep under-ground, under-sea or
under-ice. These hot topics include the search for dark matter, the
Majorana nature of neutrinos, the solar neutrinos or other
astrophysical neutrino sources. We have seen in this conference
important advances in several of these topics. In dark matter searches,
for intance, we are in the very thrilling situation in which there are
hints of first detection together with exclusion limits which in their
simplest interpretation seem incompatible with such signals. In the
mean time, we are likely attending the dawn of the neutrino
astronomy. The events observed by IceCube are very difficult to
explain from the background due to atmospheric muon and neutrinos, not
only for the observed rates at high energies, but also for the ratio
between showers and tracks, the observed spectral index or the
distribution of the vertices in the detector. This first evidence of
cosmic neutrinos should be confirmed by more data from IceCube
and by the future detectors like KM3NeT and Baikal-GVD, already
successfully deploying the first prototype lines. In paralel, we have
also seen the potential of the multi-messenger approach, which allow
to increase the sensitivy using the information of other channels like
gamma rays or gravitational waves. Next years will provide us many
more information (and probably puzzles) with all the data we will gather under
the surface.

\section*{Acknowledgements}
The authors acknowledge the financial support of the Spanish Ministerio de Ciencia e Innovaci\'on (MICINN), grants FPA2009-13983-C02-01, FPA2012-37528-C02-01, ACI2009-1020, Consolider MultiDark CSD2009-00064 and of the Generalitat Valenciana, Prometeo/2009/026.

\end{document}